\documentclass[aip,jcp,graphicx,reprint,flotfix]{revtex4-1}

\usepackage{inputenc}
\inputencoding{latin1}
\usepackage[T1]{fontenc}
\usepackage{graphicx}
\usepackage{palatino} 
\usepackage{mathpazo} 
\usepackage{bm} 
\usepackage{color}
\usepackage[usenames,dvipsnames]{xcolor}

\usepackage{hyperref}

\begin{document}

\title{High pressure specific heat spectroscopy reveals 
  simple relaxation behavior of glass forming molecular liquid}

\author{Lisa Anita Roed} \author{Kristine Niss} \author{Bo Jakobsen}
\email{boj@dirac.ruc.dk} 
\affiliation{DNRF Centre ``Glass and Time'',
  IMFUFA, Department of Sciences, Roskilde University, Postbox 260,
  DK-4000 Roskilde, Denmark }


\begin{abstract}
  The frequency dependent specific heat has been measured under
  pressure for the molecular glass forming liquid 5-polyphenyl-4-ether
  in the viscous regime close to the glass transition. The temperature
  and pressure dependence of the characteristic timescale associated
  with the specific heat is compared to the equivalent timescale from
  dielectric spectroscopy performed under identical conditions. It is
  shown that the ratio between the two timescales is independent of
  both temperature and pressure. This observation is 
  non-trivial and demonstrates the existence of specially simple
  molecular liquids in which different physical relaxation processes
  are both as function of temperature and pressure/density governed by
  the same underlying ``inner clock''. Furthermore, the results are
  discussed in terms of the recent conjecture that van der Waals
  liquids, like the measured liquid, comply to the isomorph theory.
\end{abstract}

\maketitle{}
The viscosity of liquids close to the glass transition is strongly
temperature dependent -- just a few percent decrease in temperature
can lead to several decades increase in viscosity. Coupled to the
increase in viscosity is a slowing down of the molecular structural
relaxation time (the alpha relaxation time), leading to a timescale
separation between the fast iso-structural degrees of freedom and the
slow structural degrees of freedom.  The glass transition occurs at
the temperature where the slow degrees of freedom are no longer
accessible on the experimental timescale. This gives rise to a drop in
the measured specific heat which is a classical signature of the glass
transition \cite{kauzmann1948}. The time scale separation in the
highly viscous liquid just above the glass transition temperature
leads to time or equivalently frequency dependence of physical
properties which couple to the structural relaxation; the mechanical
moduli and the dielectric constant are for example frequency
dependent. Even though heat capacity is a classical probe of the glass
transition, the awareness of the fact that the specific heat is also a
frequency dependent response function which shows relaxation is
surprisingly young and only dates back to the 80'es (some of the
earliest discussions are given in
Refs.\ \onlinecite{Gobrecht1971,Angell1983,birge1985,christensen1985,Birge1986,Nielsen1996}).
The first experimental specific heat spectroscopic techniques for
studying viscous liquids were also developed in the 80'es by Birge and
Nagel \cite{birge1985} and Christensen \cite{christensen1985}. The
amount of data on frequency dependent specific heat is still very
limited today 30 years later, probably because no standard commercial
technique has been available. (The following list is not a
comprehensive list of all specific heat spectroscopy studies of glass
forming liquids, but covers to the best of our knowledge all groups
that have addressed the issue, Refs.\ 
\onlinecite{christensen1985,birge1985,Settles1992,Moon1996,christensen1998,Carpentier2002,Bentefour2003,Minakov2003,jakobsen2010}).

The workhorse in the study of frequency dependent response of glass
forming liquids is dielectric spectroscopy both when it comes to
studies of the temperature dependence of the alpha relaxation (e.g.\
Refs.\ \onlinecite{Richert1998,Hecksher2008,Elmatad2009}) and the
spectral shape of the relaxation (e.i.\ stretching
\cite{Bohmer1993,Albena09} and beta-relaxation \cite{Johari1970}). The
use of dielectric spectroscopy is particularly dominant when it comes
to high pressure studies, which have become increasingly important
during the last couple of decades. Today it is clear that the dynamics
of viscous liquids should be understood as function of both
temperature and pressure/density because this is the only way to
disentangle the effect of density from that of thermal energy. Two key
findings from high pressure studies are: Density scaling (e.g.\ Refs.\
\onlinecite{dreyfus2003,alba2004,casalini2004,Roland2005} ), and
isochronal superposition (e.g.\ Refs.\
\onlinecite{tolle01,roland2003,pawlus2003,ngai2005,roland2008,roed2013}).
These results are almost exclusively based on dielectric data because
dielectric spectroscopy is easily adapted to high pressure and
commercial equipment is available \footnote{Examples of commercial
  high pressure dielectric equipment are a full dielectric high
  pressure system from ``NOVOCONTROL Technologies GmbH'' and high
  pressure vessels with electrical feed throughs from ``Unipress
  Equipment''.}.

Different physical properties probe the microscopic dynamics in
different ways. An example is that dielectric spectroscopy only is
sensitive to degrees of freedom which involve reorientation of dipoles
in the liquid while specific heat measures those degrees of freedom
which couple to changes in energy. This naturally leads to differences
in the observed characteristic time scales for different properties it
is e.g.\ well-known that shear-mechanical relaxation is faster than
dielectric relaxation (e.g.\ Ref.\ \onlinecite{jakobsen2005} and references
therein).

While all the liquid dynamics slows down upon cooling (or compression)
it is by no means trivial that all time scales follow each other as a
function of temperature and pressure. A well-known example is the
pronounced decoupling between translational and rotational motions
which has been confirmed in many systems, also under pressure (e.g.\
Ref.\ \onlinecite{Bielowka2001}). Even time scales which at a first
coarse look seem to follow each other over many decades, have been
shown to have differences in the temperature dependence when analyzed
in detail (e.g.\ Refs.\
\onlinecite{Stickel1996,jakobsen2005}). Moreover, there is an
increasing amount of evidence for dynamical heterogenities in viscous
liquids (e.g.\ Ref.\ \onlinecite{Ediger2000}). Different physical
properties will a priori ``see'' and average differently over
the dynamical heterogeneities. This would imply different
temperature-dependence of the timescales as the dynamical
heterogenities evolve with cooling and lead to the picture suggested
by Angell in 1991, namely a series of decoupling temperatures as the
liquid is cooled down\cite{Angell1991}.

An example of the pitfall of exclusively basing analysis on one
response function is the understanding of the dynamics in monohydroxy
alcohols. It was for a long time thought that the main dielectric
relaxation peak was the signature of the structural relaxation,
leading to puzzling observations about monohydroxy alcohols (e.g.\ in
Ref.\ \onlinecite{Bohmer1993}). This was only resolved when other
response functions were analyzed, showing that the dominant process in
the dielectric spectrum is not related to the structural molecular
relaxation \cite{Bohmer2014} seen in specific heat \cite{Huth2007} and
shear modulus \cite{Jakobsen2008,Hecksher2014}.

It is in itself a fundamental question whether the liquid relaxation
seen in different techniques behaves in the same way. Additionally, it
is important to establish whether dielectric results especially at
elevated pressures can be generalized and viewed as generic
information about the alpha relaxation.

Experimentalists have often attempted to find general (universal)
behaviors and correlations in the effort to guide theory and models
for the viscous slowing down
\cite{Bohmer1993,sokolov93,scopigno03,novikov04}. However, as more and
more systems are studied, these results are usually found to hold only
for a limited class of systems
\cite{yannopoulos00,huang01,buchenau04a,Niss2007,Dalle-Ferrier10}. Another
trend is to focus on exotic phenomena seen in complicated systems like
the notorious counter example water
\cite{Debenedetti2003,Angell2008}. The emerging picture is that while
the viscous slowing down and the glass transition as such are
universal features which (at least in principle) can be observed in
all systems independent of chemical details, there is also a myriad of
specific behaviors and it seems unlikely to capture all this in one
simple model. Based on this understanding our proposal is to address
the question of what the simplest behavior is? Or put in other words
what are the features that should be included in the ``ideal gas
model'' of glass forming liquids? 

In this work we present frequency dependent specific heat measurements
taken as function of both temperature and pressure up to 300 MPa on
the molecular glass former 5-polyphenyl-4-ether (\textit{5PPE}) and
compare to existing dielectric data \cite{roed2012,roed2013} taken in
the same pressure equipment ensuring consistency of absolute
temperature and pressure. 
To the best of our knowledge these are the first ever high pressure
data  where both dielectric and specific heat spectroscopy are
taken under identical conditions, allowing for a detailed
comparison. 

5PPE has been studied intensively by the ``Glass and time'' group over
the past years (at atmospheric pressure
\cite{niss2005,jakobsen2005,hecksher2010,jakobsen2010,jakobsen2012,hecksher2013},
and at high pressures \cite{gundermann2012,roed2013,xiao2015}), and
has been found to have a particular simple behavior.  One of these
findings was that no decoupling is seen in 5PPE when comparing the
temperature dependence of 6 different timescales
\cite{jakobsen2012}. This result was later supported by result from
the group of C. Schick who
compared dielectric and specific heat spectroscopy with lower accuracy
but over larger range in dynamics \cite{Shoifet2015}.

The data in the current paper extends these studies to high pressures. 
Hereby addressing the question of whether the alpha relaxation time is
uniquely determined independent of probe in the entire phase space.

The strong temperature dependence of the characteristic timescales of
liquids close to the glass transition, has the consequence that even
small temperature/pressure differences will lead to large errors in
the measured timescales. In order to make detailed comparisons it is
therefore crucial to measure different response functions under the
same conditions (see e.g.\ Ref.\ \onlinecite{Richert1992}). To the
best of our knowledge there is only one previous study of frequency
dependent specific heat at elevated pressures, in which Leyser
\textit{et al.}\ \cite{Leyser1995} 20 years ago investigated
orthoterphenyl in a limited pressure range up to 105 MPa, and compared
to existing literature data \cite{Naoki1987}. Their data show that the
timescales do follow rather closely along the investigated paths in
phase space, however some systematic changes are seen. Due to the
limited pressure/temperature range investigated and the fact that the
dielectric data they compare to is from a different study, it is not
possible to conclude if the ratio between the two timescales is
constant or changing for this liquid.

Our unique specific heat spectroscopy technique is based on measuring
the thermal impedance at a spherical surface in the liquid (the
relation between heat flow into the liquid from the surface, and the
temperature at the same surface). The method is inspired by the method
developed by Birge and Nagel 30 years ago
\cite{birge1985,Birge1997}. Our version of the technique
\cite{jakobsen2010} utilize a spherical geometry, and a liquid layer
much thicker than the thermal wavelength (the thermally thick
limit). In this case the outer thermal and mechanical boundaries do
not influence the measured property \cite{christensen2008}. What is
measured is the longitudinal volume specific heat $c_{l}$
\cite{christensen2007,christensen2008} which is approximately equal to
the isobaric specific heat \footnote{The longitudinal volume specific
  heat $c_{l}$ is related to the isochoric specific heat ($c_{v}$) by
  $c_{l}=\frac{K_{S}+\frac{4}{3} G}{K_{T}+\frac{4}{3}G}c_{v}$ where $K_{S}$ and
  $K_{T}$ respectively are the adiabatic and the isothermal bulk moduli,
  and $G$ is the shear modulus. At state points where relaxation
  occurs $G$ will differ from 0, which means that $c_{l}$ and $c_{p}$
  differs. However, the difference in characteristic timescale is very
  small (see supplementary material of \onlinecite{jakobsen2012}).}. By
the 3omega technique a temperature dependent resistor is utilized
as heat generator and thermometer simultaneous
\cite{Birge1997,jakobsen2010}.

The method is well suited to be adopted to different sample
environments with little requirements to the mechanical properties of
the sample cell and electrical connections. Altogether this makes the
method perfect for integration in existing pressure equipment.

The measurements were performed using commercial high-pressure
equipment from Unipress Equipment (Warsaw, Poland).  The pressure is
applied using a pressure liquid, which is separated from the sample
cell by a shielding of rubber and Teflon. Pressures go up to 600 MPa
with a stability of 3 MPa, temperatures ranges from 233 to 333 K (for
further details see Refs.\ \onlinecite{Igarashi2008b,gundermann2012,roed2012}).
 A spherical NTC-thermistor bead (a temperature dependent
resistor, with ``Negative Temperature Coefficient'') is used as heat
generator and thermometer and is placed in the middle of the
sample cell
with a distance of $\approx 10$ mm from the closest sides ensuring approximately
thermally thick conditions down to the millihertz range.
The 5PPE liquid studied is the diffusion pump oil 5-polyphenyl-4-ether
acquired from Santovac. The liquid was dried for one hour under vacuum
before use.

\begin{figure}
  \includegraphics{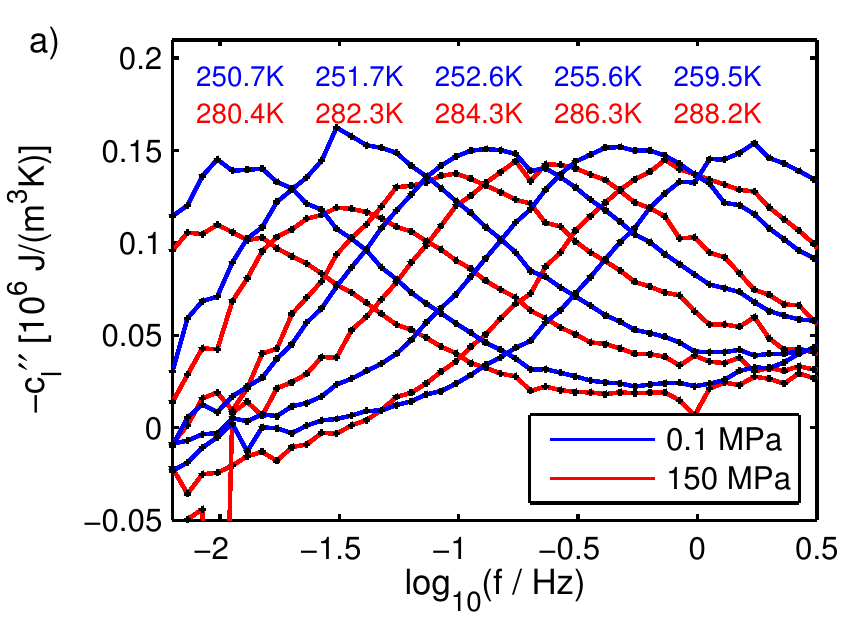}
  \includegraphics{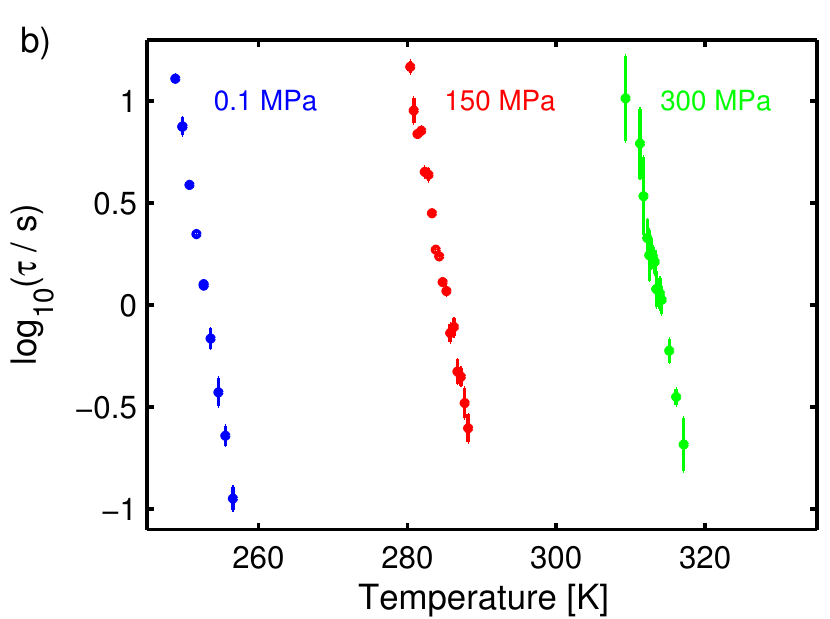}
  \caption{Specific heat data. a) The imaginary part of the
    longitudinal specific heat at the indicated temperatures and
    pressures. b) The relaxation time $\tau$, based on the loss-peak
    frequency, as a function of temperature for all studied
    temperatures and pressures. }
  \label{figure1}
\end{figure}

The specific heat measurements are performed at different temperatures
along the isobars; 0.1 MPa (atmospheric pressure), 150 MPa, and 300
MPa. The initial data analysis was performed as in Ref.\
\onlinecite{jakobsen2010}.  Examples of the imaginary part of the
frequency dependent longitudinal specific heat at different
temperatures and pressures are shown in Fig.\ \ref{figure1}a. The peak
frequency of the imaginary part, $f_{{c_{l},{\mathrm{lp}}}}$ (the
loss-peak frequency) defines a directly experimental accessible
characteristic timescale ($\tau_{c_{l}}=1/(2\pi
f_{{c_{l},{\mathrm{lp}}}})$) for the specific heat relaxation. Figure
\ref{figure1}b. shows the relaxation times for all measured
temperatures and pressures. The extreme temperature and pressure
dependence of the relaxation time is clearly seen in this
representation. The shown error bars refer to limitations in
  precision of the specific heat technique \cite{jakobsen2010}. There
  is moreover approximately 0.1 decade uncertainty on the relaxation time
  due to the limited stability of the pressure of 3MPa. 

\begin{figure}
 \includegraphics{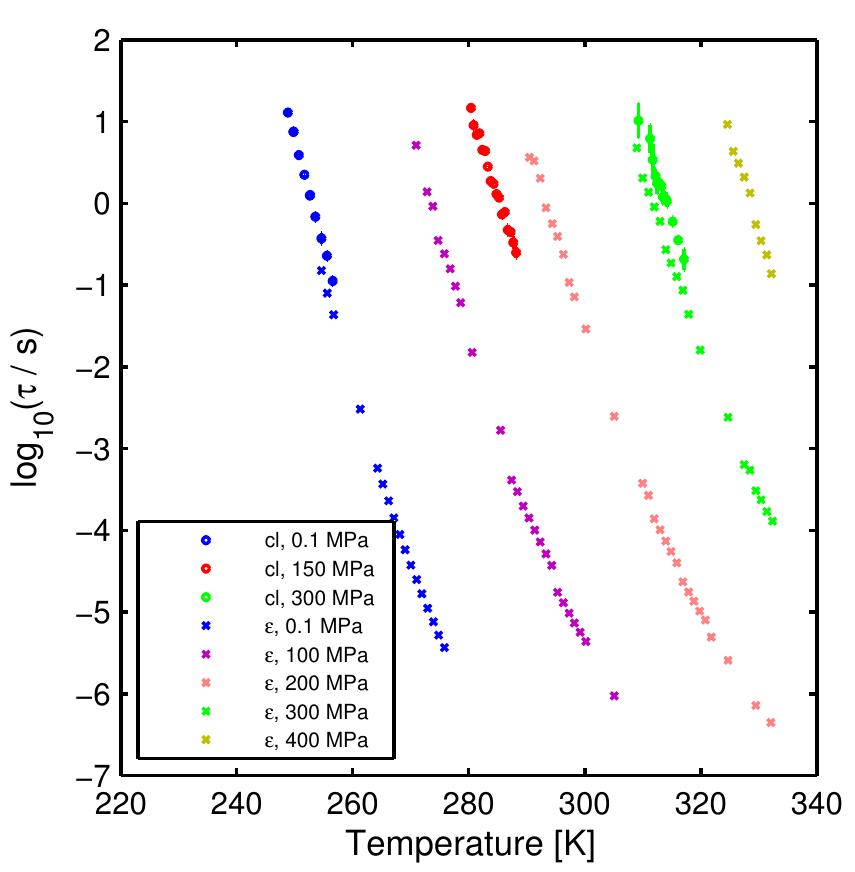}
  \caption{Comparison of the relaxation times ($\tau$) for the
    specific heat measurements (same data as on Fig. \ref{figure1})
    and the dielectric measurements at the indicated pressures.}
  \label{figure2}
\end{figure}

The comparable dielectric spectroscopy measurements were taken in
relation to Refs.\ \onlinecite{roed2013,roed2012} utilizing the same
pressure equipment. The dielectric measurements were performed along
the isobars; 0.1 MPa, 100 MPa, 200 MPa, 300 MPa, and 400 MPa. The
relaxation time is again defined from the loss-peak frequency
($\tau_{\epsilon}=1/(2\pi f_{{\epsilon,{\mathrm{lp}}}})$). Figure
\ref{figure2} shows the relaxation time as a function of temperature
at the different pressures for both the specific heat and for the
dielectric measurements. As mentioned earlier the fact that the data
are taken under the same thermal and pressure conditions gives a
unique possibility for directly comparing the temperature and pressure
dependence of the timescale from dielectric and specific heat
spectroscopy over a rather wide pressure range.

From Fig.\ \ref{figure2} it is seen that the temperature dependence of
the relaxation times from the two methods closely follows each other
at all investigated pressures. It is also seen that the dielectric
relaxation time is faster than the specific heat relaxation time at
all pressures.  This shows that the ambient pressure observation of
identical temperature dependence of dielectric and specific heat
relaxation time presented by some of us in Jakobsen \textit{et al.}
(2012) \cite{jakobsen2012} on 5PPE 
also holds at elevated pressures.

\begin{figure}
  \includegraphics{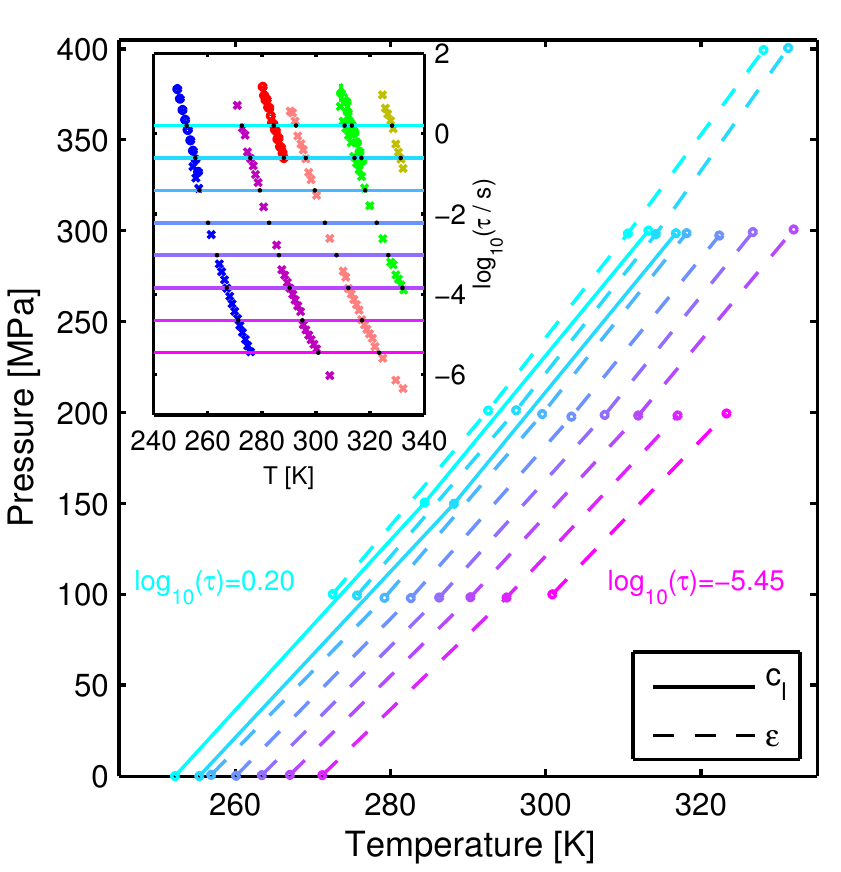}
  \caption{Isochrones for specific heat (full lines) and dielectric
    (dashed lines). The two slowest isochrones are found for both the
    specific heat and dielectric, but the six fastest isochrones are
    only found for the dielectric measurements. The inset show the
    same data as Fig.\ \ref{figure2} where the vertical lines
    illustrate eight chosen relaxation times. The temperatures used
    for the isochrones are found by interpolation (indicated by small
    solid points on the insert).}
  \label{figure4}
\end{figure}

Figure \ref{figure4} presents the same data using isochrones, which
are lines in the phase diagram of constant relaxation time; that is
contour lines of the relaxation time map, $\tau(T,P)$. The isochrones
are illustrated for both methods, and it is seen that the isochrones
for the two response functions are parallel. 

The \textit{main result} of this article  is that the
$\tau_\epsilon/\tau_{c_{l}}$ ratio is constant within error bars for
all investigated temperatures and pressures for 5PPE. This result is
obtained by combining the observation from Fig.\ \ref{figure2} (and
Ref.\ \onlinecite{jakobsen2012}) and Fig.\ \ref{figure4} as described
below.  Combining the data from the present study and from
  Ref.\ \onlinecite{jakobsen2012} yields 
  $\log_{10}(\tau_\epsilon/\tau_{c_{l})}=-0.4\pm0.1$, with the major
  contribution to the uncertainty coming from pressure instabilities.

The relaxation time is a smooth function of temperature and pressure,
and parallel isochrones therefore imply that an isochrone for one of
the two response functions is also an isochrone for the other but with
a different timescale, as the dielectric relaxation is faster than
the specific heat relaxation at a given state point.

The difference between the timescales associated with the isochrones
is not \textit{a priory} the same for different isochrones. The
observation that the $\tau_\epsilon/\tau_{c_{l}}$ ratio is constant
along the isobars as shown on Fig.\ \ref{figure2} (and illustrated
even more clearly at atmospheric pressure in Ref.\
\onlinecite{jakobsen2012}) is therefore an additional simplicity. Since any
isobar crosses all the isochrones, we can conclude that the timescale
difference associated with the isochrones indeed is the same in all of
the explored part of the phase diagram for this liquid (it is enough
to shown that it is constant along one curve, crossing the
isochrones).

The observation that the isochrones from one of the methods are also
isochrones for the other method is a direct prediction of the isomorph
theory developed in the ``Glass and Time'' group
\cite{gnan2009,Dyre2014,Schroder2014}. The theory predicts the existence of
isomorphs which are curves in the phase diagram of simple (called
\textit{Roskilde simple}) liquids along which a number of properties
are invariant.  The isochrones of any response function will in the
viscous regime (or in reduced units) coincide with an isomorphs, which
means that an isochrone from one response function is also an
isochrone for another response function, as all dynamics are invariant
along an isomorph \cite{roed2013}.

Simple van der Waals liquids as 5PPE are expected to be Roskilde
simple \cite{gnan2009,Ingebrigtsen2012,Dyre2014}. We have in Refs.\
\onlinecite{gundermann2012,xiao2015,roed2013} shown that 5PPE corroborates
other isomorph predictions, this together with the present result
indicating that 5PPE indeed is a good example of a Roskilde simple
liquid.

In Jakobsen \textit{et al.}\ (2012)\cite{jakobsen2012} some of us showed
that at ambient pressure, not only specific heat and dielectric, but
also several other thermo-viscoelastic responses have timescales which
follow closely with temperature for 5PPE. Based on this we hypothesize
that the timescales of all the frequency dependent response functions
of 5PPE have a common temperature and pressure dependence in the
viscous regime. An interpretation of this observation is that the
structural relaxation time is governed by one ``inner clock'' and
implies a greater simplicity than predicted by the isomorph theory.
It is also a challenge to understand how all response functions can
couple so closely if the underlying dynamics is heterogeneous with a
growing length scale along with viscous slowing down --- since the
different macroscopic responses could very well average differently
over the heterogeneities.

5PPE has other simple behaviors which are beyond isomorph theory. The
spectral shape of the dielectric signal obeys Time Temperature
Pressure Superposition (TTPS), thus it is independent of pressure and
temperature in a range where $\tau_{\epsilon}$ changes by 7 orders of
magnitude \cite{roed2012,roed2013}. Time Temperature Superposition is
moreover found in frequency dependent bulk modulus and the shear
mechanical relaxation, and these two moduli in fact have the same
spectral shape \cite{hecksher2013}.  Finally 5PPE is also found to
obey Time Aging Time Superposition \cite{hecksher2010}.

The simple behavior of 5PPE is obviously not universal for all
glass-forming liquids, it might not even be a general behavior of van
der Waals bonded liquids. Yet 5PPE exhibits the hallmarks of glass
forming liquids; it has a non-Arrhenius temperature dependence
(fragility index m=80--85)\cite{MasterThesis} and a non-exponential
relaxation (high frequency power law -0.5)\cite{jakobsen2005}. This
means that models and theories for understanding non-Arrhenius
non-exponential relaxation need to be consistent with a simple
behavior where there is no decoupling of different timescales and no
increase in the broadening of the relaxations in the entire viscous
range (defined as timescales ranging from a microsecond up to a kilosecond).

To summarize, we have presented specific heat spectroscopy data over
an unprecedented pressure range, with accompanying dielectric data
taken under the same thermal and pressure conditions. The main
experimental results are that the characteristic timescale of the
specific heat and the dielectric relaxation follow each other closely
as function of temperature at all investigated pressures, and that
isochrone curves for specific heat and dielectric spectroscopy
coincide. The consequence of these two observations is that the
$\tau_{\epsilon}/\tau_{c_{l}}$ ratio is constant over the investigated
temperature and pressure range, with the dielectric spectroscopy being
the fastest.

Altogether 5PPE seems to have the simplest possible behavior in
respect to differences in timescales between response functions,
namely that one common inner clock controls the different relaxations,
and the only difference is a temperature and pressure independent
ratio between the different
timescales.

\begin{acknowledgments}
L.\ A.\ R.\ and K.\ N.\ wish to acknowledge The Danish Council for
Independent Research for supporting this work. The center for viscous
liquid dynamics ``Glass and Time'' is sponsored by the Danish National
Research Foundation's Grant No. DNRF61.
\end{acknowledgments}


%

\end{document}